\documentclass{statsoc}
\usepackage{amsfonts}
\usepackage{amsmath}
\usepackage{graphicx}
\usepackage{natbib}
\title[Multivariate latent growth model for student careers]{A general multivariate latent growth model with \\ applications in student careers Data warehouses}
\author[Bianconcini S. ]{Silvia Bianconcini\footnote{\emph{Address for correspondence}: Silvia Bianconcini, Department of Statistical Sciences, University of Bologna, Via Belle Arti, 41 - 40126 Bologna, Italy. E-mail:silvia.bianconcini@unibo.it.}}
\address{Department of Statistical Sciences,
         University of Bologna,
         Italy.}
\email{silvia.bianconcini@unibo.it}
\author[S. Bianconcini and S. Cagnone]{Silvia Cagnone}
\address{Department of Statistical Sciences,
         University of Bologna,
         Italy.}

\keywords{University evaluation, student capability, Data warehouse,
longitudinal and mixed data, generalized linear latent variable
models.}

\begin{document}

\begin{abstract}
The evaluation of the formative process in the University system has
been assuming an ever increasing importance in the European
countries. Within this context the analysis of student performance
and capabilities plays a fundamental role. In this work we propose a
multivariate latent growth model for studying the performances of a
cohort of students of the University of Bologna. The model proposed
is innovative since it is composed by: (1) multivariate growth
models that allow to capture the different dynamics of student
performance indicators over time and (2) a factor model that allows
to measure the general latent student capability. The flexibility of
the model proposed allows its applications in several fields such as
socio-economic settings in which personal behaviours are studied by
using panel data.
\end{abstract}

%\doublespacing

\section{Introduction}

The Bologna Process started in 1999 with the aim of creating a
European Higher Education Area, in which students could choose from
a wide and transparent range of high quality courses and benefit
from smooth recognition procedures. The Bologna Declaration has
initiated a series of reforms needed to make European Higher
Education more compatible and comparable, more competitive and more
attractive than before, both for Europeans and for students from
other continents. Hence, the evaluation of formative processes has
received a growing attention by policy makers
 and public agents in order to identify critical
 factors for achievements that can improve curricula, instructional strategies, and conditions for learning.
 \\An important emerging problem is the comparison between students' performances \textit{i)}
 when different supporting and tutoring actions are adopted during the course of studies,
 \textit{ii)} in presence of different personal situations.
With this purpose, several Universities have created Data WareHouse
(DWH) systems to  collect detailed multivariate individual responses
over time, which consist of  mixtures of count, categorical, and
continuous observations. These longitudinal data allow to answer
questions about student progress, evaluate how each individual
performs over time (\emph{within-individual change}), predict the
main differences among individuals in their change
(\emph{interindividual differences in change}). However, in presence
of multidimensional observations, a challenging problem is the
characterization of both temporal and cross-sectional dependencies
among response variables having different measurement scales. In
such cases, it is natural to consider models in which the dependency
among the responses is due to the presence of both one or more
latent variables and random effects, as shown in several approaches
developed in the literature.\\ \cite{RoLi:00} proposed a 2-step
linear mixed model applied to multiple continuous outcomes.
 These authors use time-dependent factors to account for correlations of items within time.
 On the other hand, a random intercept and random effects are introduced for explaining
 the correlations across time of both items and time-dependent latent variables, respectively.
An extension to such models is provided by \cite{Du:03}, who
introduced a dynamic latent trait model for multidimensional
 longitudinal data
in the context of the Generalized Linear Latent Variable Model
(GLLVM) so that different kinds of observed variables can be
considered. An autoregressive structure that allows for covariates
is used to model the structural part.  The model is estimated by
using the MCMC procedure. Within the same framework, a full
information likelihood estimation method via the EM algorithm
is developed by \cite{CaMo:09} with particular attention to ordinal data.\\
A very general approach is represented by multilevel models
\citep{SkRa:04} that allow to deal with longitudinal and/or
multidimensional mixed data. In presence of repeated measures,
occasions are viewed as first level units whereas respondents are
second level units. With multidimensional data, first level units
are represented by items nested within individuals. When both the
dimensions are considered more complex hierarchical structures
 have to be taken into account.\\
Multidimensional and longitudinal data are also treated within the traditional structural
equation approach (SEM). They are modeled in two different ways.\\
According to the first one \citep{JoSo:01}, a standard confirmatory
factor model is considered and its the main feature is that the
corresponding error terms are correlated over time. Moreover, the
latent variables are identified by setting to 1 the same loading
over time.\\
The second approach is represented by latent growth models, widely
applied in the analysis of change \citep{SiWi:04}. Random effects
are included into the model to account for the individual
differences both in the initial status and in the rate of growth.
The peculiar feature of such models is that the random coefficients
are treated as latent variables within the traditional SEM approach
\citep{MuKh:98}. In this context, univariate analysis are usually
performed by studying the temporal dynamics of a single indicator,
considered as a proxy of the individual performance. Multivariate
extensions essentially consist of modeling the trajectories of
several items separately, and then to allow for correlations among random coefficients \citep{BoCu:06,Ra:07}.\\
This work is motivated by the study of the data coming from the DWH
of the University of Bologna. We focus on the achievements of a
cohort of students enrolled in 2001 at the Faculty of Economics.
Multiple items are present in the data set. Their behaviour over
time can be classically analyzed by means of multivariate latent
curves, where all the variability between items is captured by the
correlation of the random coefficients. However, some of these items
can be seen as indicators of student latent capabilities. Hence,
part of their variability can be due to latent constructs. In order
to take into account these two important aspects simultaneously we
propose a new, general, class of models, that consists of two parts:
\textit{i)} multivariate latent curves that describe the behaviour
of each item over time, \textit{ii)} a factor model that specifies
the relationship between manifest and latent variables. Although
these two components have been widely developed in the literature
separately, the novelty of our proposal lies in integrating them
into a unique framework.

 The model is developed within the GLLVM
framework. According to this approach, the response variables are
assumed to follow different distributions of the exponential family,
with item-specific linear predictors depending on both time-specific
covariates,
  latent variables, and measurement errors.
  Moreover, we extend the GLLVM by including item-specific random coefficients so that each item has its own trajectory over
  time.
  Our approach has clear advantages with respect to both multivariate latent curves since the unobservable capabilities
  of the single individual can be taken into account, and the dynamic factor
  models since we incorporate a more flexible treatment of the temporal dynamics of
the items.
  The latter are also assumed to be heteroscedastic over time.  \\
The paper is organized as follows.
 In Section 2 we present the data source and perform an exploratory analysis to demonstrate the potential
 of our approach in describing student performances over time.
 Section 3 describes the proposed methodology in terms of model specification, identification and estimation.
  In Section 4 we present the results of model estimation for the overall
 data set and for different temporal patterns observed in the sample.
 We conclude with a discussion in Section 5.

\section{Data}

The data set analyzed was extracted from the Data warehouse of the
University of Bologna. This latter is a system that collects and
constantly updates informations by integrating data coming from
sources of different nature. The project started in 2002 in order to
support planning, control and decision processes. \par The DWH
contains a great amount of information per each student and allows
to build the overall university student career. It is also possible
to find socio-demographic information (gender, country/region of
origin, etc.) and the mark obtained in the final exam of the High
School. We decided to analyze the cohort of $n = 821$ students
enrolled at the Faculty of Economics in the academic year 2001/2002
since this Faculty is one of the biggest of the University of
Bologna and such year is the first available in the DWH, so that
several time points can be observed.
\begin{figure}[!htbp] \centering
\makebox{\includegraphics[height=4.5cm,width=4.5cm]{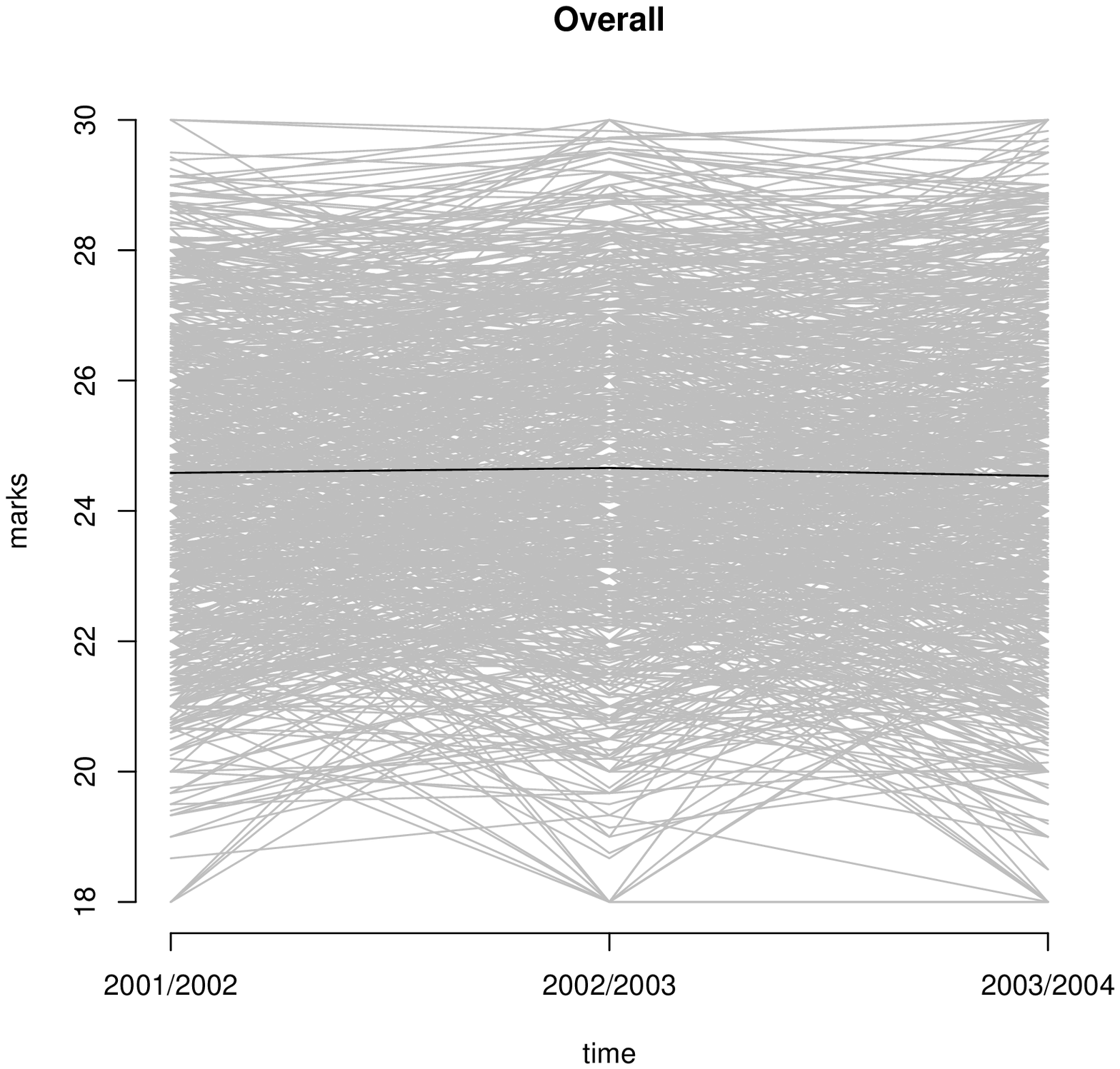} %\hspace{-1.5pt}
\includegraphics[height=4.5cm,width=4.5cm]{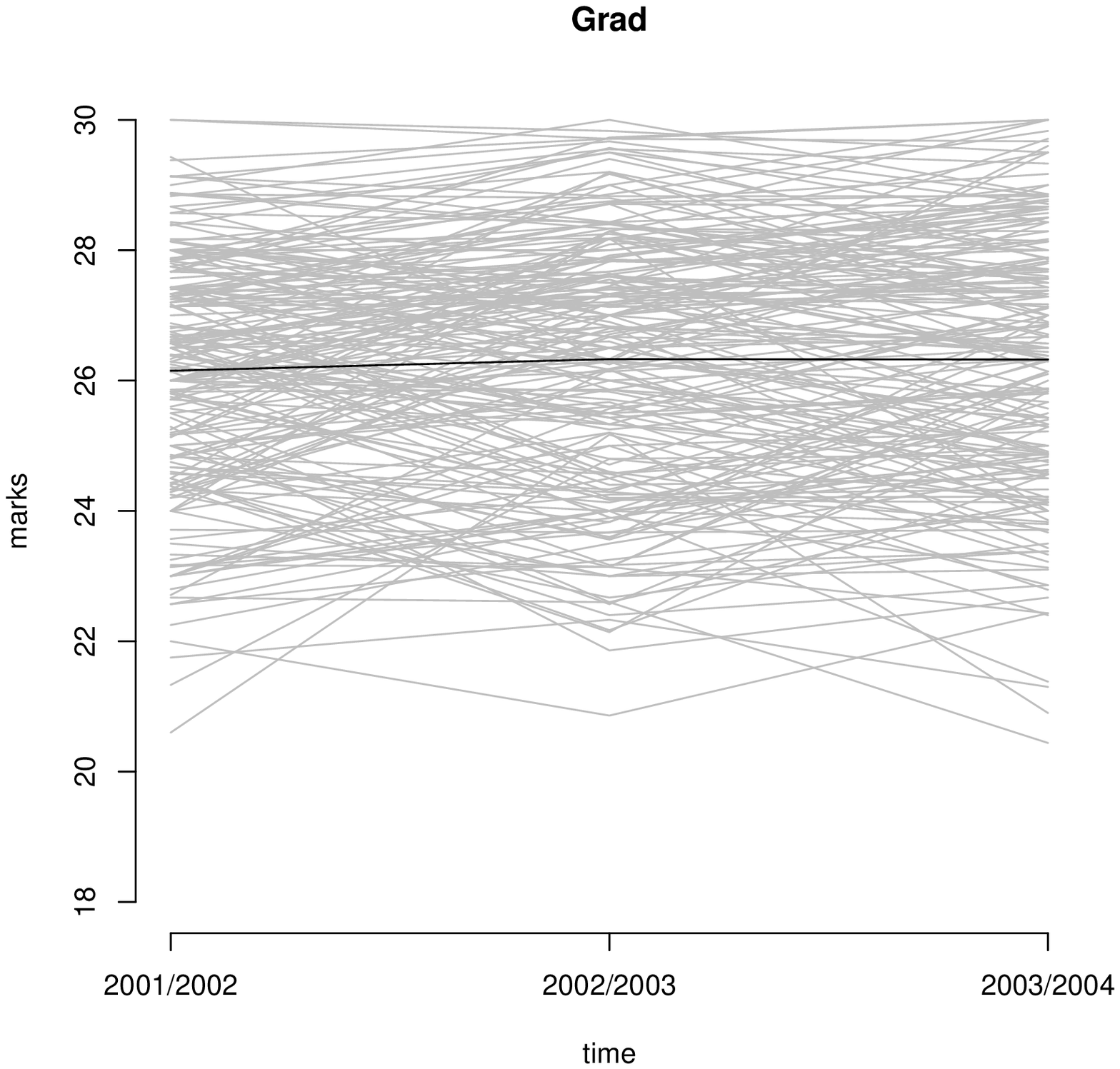}%}\hspace{-1.5pt}
\includegraphics[height=4.5cm,width=4.5cm]{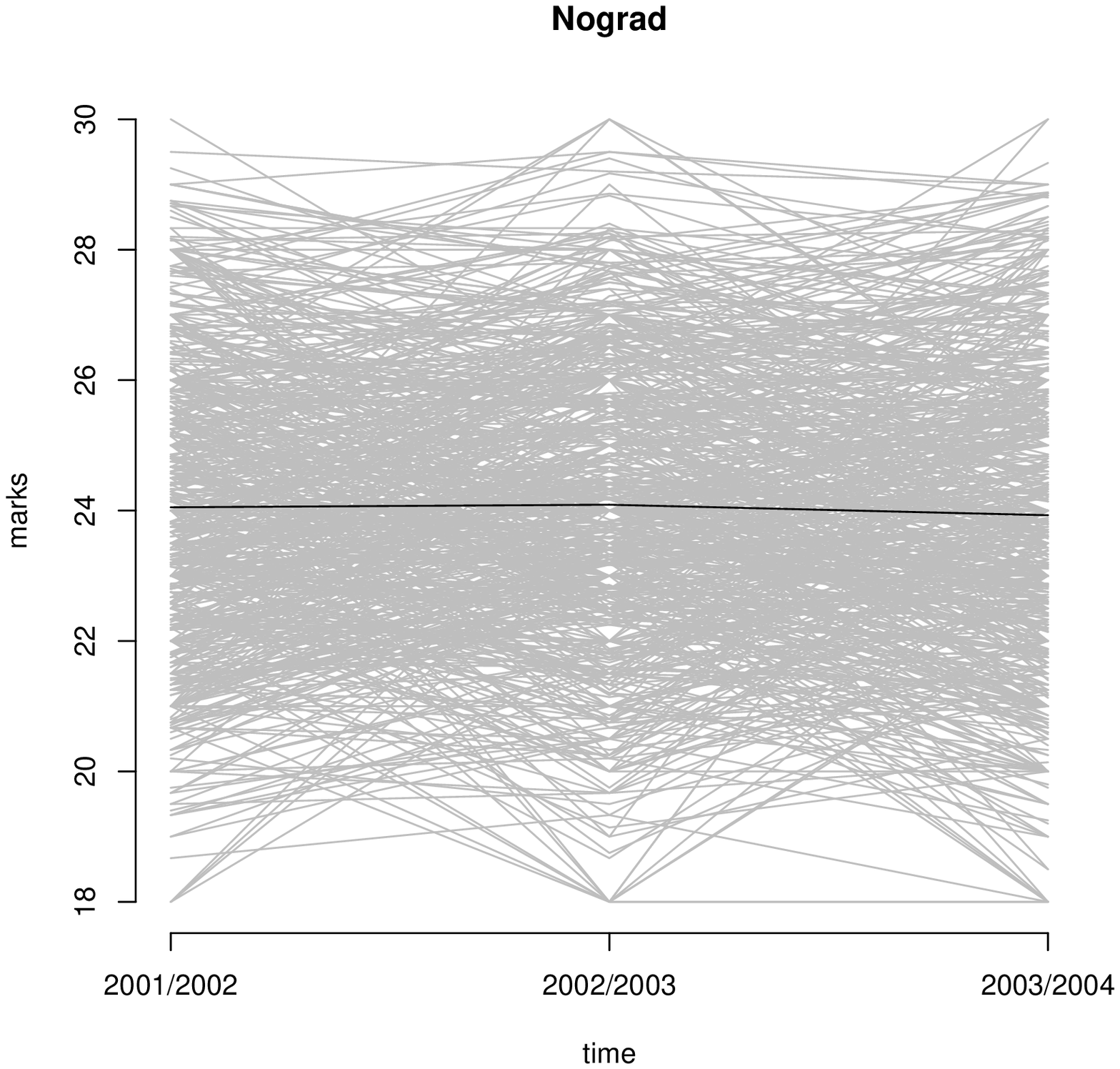}}
\caption{\label{fig:1} Trajectories of AM in the three academic
years for Overall, Grad, Nograd patterns.}
\end{figure}
Therefore, this data set does
not contain missing data in the first three years of the study.
After the third year, the presence of missing data is due to
different reasons: either students that got the degree in time
(three years), or drop outs, or simply missing information. Hence,
we analyzed the performance of the selected students in the first
three academic years: $t_{1} = 2001/2002$; $t_{2} = 2002/2003$;
$t_{3} = 2003/2004$. In the data set,  two variables are available
for evaluating the student performance over time: the marks obtained
in every exam  and the number of exams taken per each time point.
The average of the marks (AM) per each student in each academic year
is considered.
%\end{center}
In the left side of Figure \ref{fig:1} the AM trajectories for the
full sample of students are reported. They vary from 18 (minimum
mark to pass the exam) to 30 (maximum mark) even if they are mostly
concentrated in the range 21-28, and the overall mean (black line)
is over 24 for all the observed time points. Within the selected
sample we can distinguish two different temporal behaviours: the
first one concerns students who got the degree in the first three
years (Grad) while the second one concerns students that at $t_{3}$
did not manage to get the degree yet (Nograd). Indeed, as shown in
the center and in the right side of Figure \ref{fig:1}, the former
presents a higher overall average mark and a lower variability than
the latter. In Table \ref{Tab:1} the descriptive statistics of AM
for the overall sample, Grad ($n_{1}=195$) and Nograd ($n_{2}=626$)
are reported. Grad presents the highest correlations over the three
time points. On the other hand, Nograd is very similar to the
Overall sample in terms of both first and second order moments.
\begin{table}
\caption{\label{Tab:1} Descriptive statistics of AM in the three
academic years}%\hspace{-4.5cm}
\centering
\fbox{\begin{footnotesize}
\begin{tabular}{l c c c c c c c c c c c c}\hline
&\multicolumn{3}{c}{\textsc{\emph{Overall}}}&&\multicolumn{3}{c}{\textsc{\emph{Grad}}}&&\multicolumn{3}{c}{\textsc{\emph{Nograd}}}\\
\cline{2-4} \cline{6-8} \cline{10-12}
&$t_{1}$&$t_{2}$&$t_{3}$&&$t_{1}$&$t_{2}$&$t_{3}$&&$t_{1}$&$t_{2}$&$t_{3}$ \\ \hline \\
Mean&$23.88$&$24.14$&$23.59$ &&$26.15$&$26.33$&$26.32$&&$23.17$&$23.46$&$22.74$\\
Std Dev&$4.62$&$4.10$&$5.26$&&$1.85$&$1.99$&$2.05$&&$4.99$&$4.34$&$5.65$\\
&\multicolumn{11}{c}{\emph{Correlations}}\\
$t_{2}$&$0.41$&&&&$0.68$&&&&$0.34$\\
$t_{3}$&$0.32$&$0.37$&&&$0.67$&$0.76$&&&$0.25$&$0.29$\\
 \hline
\end{tabular}
 \end{footnotesize}}
\end{table}
\begin{table}
\caption{\label{Tab:2}  Number of students per NE over time for the
Overall sample, Grad and Nograd}%\hspace{-9cm}
\centering
\fbox{\begin{footnotesize}
%\begin{center}
\begin{tabular}{l c c c c c c c c c c c c}
  \hline
  &\multicolumn{3}{c}{\textsc{\emph{Overall}}}&&\multicolumn{3}{c}{\textsc{\emph{Grad}}}&&\multicolumn{3}{c}{\textsc{\emph{Nograd}}}\\
 \cline{2-4} \cline{6-8} \cline{10-12}
NE  &$t_{1}$ &$t_{2}$ &$t_{3}$ && $t_{1}$ &$t_{2}$ &$t_{3}$
&&$t_{1}$ &$t_{2}$ &$t_{3}$ \\  \hline
 0&$22$&$16$&$30$&&$0$&$0$&$0$&&$22$&$16$&$30$\\
 1&$37$&$20$&$36$&&$0$&$0$&$0$&&$37$&$20$&$36$\\
 2&$59$&$60$&$53$&&$1$&$1$&$0$&&$58$&$59$&$53$ \\
 3&$98$&$95$&$67$&&$1$&$0$&$0$&&$97$&$95$&$67$\\
 4&$118$&$139$&$92$&&$6$&$9$&$0$&&$112$&$130$&$92$  \\
 5&$147$&$173$&$105$&&$23$&$32$&$3$&&$124$&$141$&$102$  \\
 6&$167$&$149$&$161$&&$61$&$63$&$53$&&$106$&$86$&$108$ \\
 7&$131$&$100$&$138$&&$70$&$54$&$65$&&$61$&$46$&$73$  \\
 8&$42$&$36$&$73$&&$33$&$13$&$34$&&$9$&$23$&$39$\\
 9 &-&$11$&$40$&& -&$5$&$21$&&-&$6$&$19$\\
 10&-&$8$&$21$&&-&$5$&$16$&&-&$3$&$5$\\
 11&-&$12$&$3$&&-&$12$&$2$&&-&$0$&$1$\\
 12&-&$2$&$0$&&-&$1$&$0$&&-&$1$&$0$\\
 13&-&$0$&$1$&&-&$0$&$0$&&-&$0$&$1$\\
 14&-&$0$&$1$&&-&$0$&$1$&&-&$0$&$0$\\
  \hline
\end{tabular}
%\end{center}
 \end{footnotesize}}
\end{table}
The number of exams (NE) is a count variable whose range is
different in the observed time points. Table \ref{Tab:2} shows the
number of students classified according to NE both taken in the
three time points and the groups defined before. The hyphens
indicate that in the first academic year, a student can take at most
eight exams. We can observe that, in general, Grad students present
the same behaviour over the three years, that is, they take a number
of exams greater than three, and concentrate it between six and
seven. On the contrary, Nograd students take a number of exams equal
or greater than zero, mostly concentrated between four to six.
Moreover, few students of the overall sample take more than eight
exams at $t_{2}$ (only $4,1\%$) and at $t_{3}$ (only $8,1\%$). It
can be useful to evaluate whether \textit{i)} the variable NE shows
a dependence over time and \textit{ii)} there is an association
between the variable AM and NE within the same time. To this aim,
the variable AM has been recoded for all the time points into four
classes according to the quartiles of the distribution. As for the
variable NE, the categories from 1 to 3 and categories from 9 to 14
have been collapsed to avoid the problem of sparseness that affects
these data in the extreme categories. In Table \ref{Tab:3} the
values of the Chi square tests (with associated p-values) are
reported for all the pairs of NE over time and for all the pairs of
AM and NE within time. The association between AM and NE for Grad is
not significant at time $t_{2}$, whereas all others are significant,
indicating that both the variables can be good indicators of the
student capability.
\begin{table}
\caption{\label{Tab:3} Associations between variables}
\centering
\fbox{\begin{footnotesize}
%\begin{center}
\begin{tabular}{c c c c c c c c c c c c }
  \\ Pairs &\multicolumn{3}{c}{\emph{Overall}} &&\multicolumn{3}{c}{\emph{Grad}} &&\multicolumn{3}{c}{\emph{Nograd}} \\ \hline
 & $\chi^{2}$ & $df$ &\emph{p-value}&&$\chi^{2}$ & $df$ &\emph{p-value}&&$\chi^{2}$ & $df$ &\emph{p-value}\\
\cline{2-4} \cline{6-8} \cline{10-12}
$\text{NE}_{t_{1}}$ vs $\text{NE}_{t_{2}}$ &  $209.10$ & $30$& $0.000$&& $73.64$ & $30$& $0.000$ &&  $80.73$ & $30$& $0.001$\\
$\text{NE}_{t_{1}}$ vs $\text{NE}_{t_{3}}$ &  $219.178$ & $30$& $0.000$&& $85.26$ & $20$& $0.000$&&  $89.51$ & $30$& $0.000$\\
$\text{NE}_{t_{2}}$ vs $\text{NE}_{t_{3}}$ &  $199.02$ & $36$& $0.000$&& $88.18$ & $24$& $0.000$&&  $110.74$ & $36$& $0.000$\\
$\text{AM}_{t_{1}}$ vs $\text{NE}_{t_{1}}$ &  $147.45$ & $15$& $0.000$&& $32.25$ & $15$& $0.006$&&  $90.94$ & $15$& $0.000$\\
$\text{AM}_{t_{2}}$ vs $\text{NE}_{t_{2}}$ &  $92.53 $ & $18$& $0.000$&& $21.39$ & $18$& $0.260$&&  $36.50$ & $18$& $0.006$\\
$\text{AM}_{t_{3}}$ vs $\text{NE}_{t_{3}}$ &  $137.81$ &  $18$& $0.000$&& $91.52$ &  $9$& $0.000$&&  $84.46$ & $18$& $0.001$\\
\hline
\end{tabular}
%\end{center}
 \end{footnotesize}}
\end{table}

\section{Modeling and estimation}

For a given student a record consists of number of exams (NE) and
the average marks (AM) achieved in each academic year. Hence, in
this section we present parametric growth models for mixed
observations, and provide special treatment to count and continuous
responses.

\subsection{Multivariate latent growth curves}

Suppose that $J$ items are observed for $n$ individuals at $T_{j},
j=1,...,J$, different time points. The measured outcomes for a
randomly selected individual are denoted by $\boldsymbol
y=(\boldsymbol y_{1}',...,\boldsymbol y_{j}',...,\boldsymbol
y_{J}')'$, where the elements $\boldsymbol
y_{j}=(y_{1j},...,y_{Tj})'$, $j=1,2,...,J$, consist of mixtures of
count ($j=1,...,J_{1}$) and continuous ($j=J_{1}+1,...,J$)
responses. In analyzing data of this type, a challenging problem is
the characterization of both the temporal and the cross-sectional
dependency among variables that have different measurement scales.
In such cases, it is natural to consider models in which the
dependencies are due to the presence of both several latent
variables and random effects, stacked into the vector $\boldsymbol
\eta$ \citep{CaMo:09}. The marginal distribution of the overall
responses is given by
\begin{equation} \label{model}
f(\boldsymbol y|\boldsymbol x)=\int g(\boldsymbol y|\boldsymbol
\eta,\boldsymbol x)h(\boldsymbol \eta)d\boldsymbol \eta
\end{equation}
where $g(\boldsymbol y| \boldsymbol \eta,\boldsymbol x)$ is the
conditional distribution of the responses $\boldsymbol y$ given a
set of covariates $\boldsymbol x$ and the latent variables
$\boldsymbol \eta$, and $h(\boldsymbol \eta)$ is their prior density
function. We refer to the GLLVM framework developed in
\cite{BaKn:99} for multivariate mixed responses and in \cite{Mou:03}
when covariate effects are included. We extend that framework to
allow for multivariate longitudinal  data. One of the main
assumption of the GLLVM approach is the conditional independence of
the responses (within  and over time) given the latent variables,
that is
\begin{equation} \label{cond}
g(\boldsymbol y| \boldsymbol \eta, \boldsymbol x)=\prod_{j=1}^{J}\prod_{t=1}^{T_{j}}g(y_{tj}|\boldsymbol \eta,\boldsymbol x _{tj})
\end{equation}
where $g(y_{tj}|\boldsymbol \eta , \boldsymbol x_{tj})$ is a distribution of the exponential family. For count data,
\begin{equation} \label{binomial}
g(y_{tj}|\boldsymbol \eta, \boldsymbol x_{tj})=\left(
                                \begin{array}{c}
                                  n_{t} \\
                                  y_{tj} \\
                                \end{array}
                              \right) \left(\frac{\exp(\upsilon_{tj})}{1+\exp(\upsilon_{tj})}\right)^{y_{tj}} \left(\frac{1}{1+\exp(\upsilon_{tj})}\right)^{n_{t}-y_{tj}} \qquad j=1,...,J_{1}
\end{equation}
where $n_{t}$ is the number of "trials" (or opportunities for an
event). Even if counts are generally modeled using a Poisson
distribution,  when the events being counted for a unit occur at a
constant rate in continuous time and are mutually independent, a
binomial distribution is more appropriate. Indeed we will deal with
counts corresponding  to the number of exams that the student takes
in each academic year,  given the maximum number of exams $n_{t}$
observed for that year. The counts have a binomial distribution if
the events for a unit are independent and equally probable. This
assumption is satisfied in our model by assuming the conditional
independence of the responses given the latent variables
$\boldsymbol \eta$.\\ On the other hand, for  continuous data
\begin{equation}\label{normal}
g(y_{tj}|\boldsymbol \eta, \boldsymbol
x_{tj})=\frac{1}{\sqrt{2\pi}\sigma_{tj}}\exp\left(-\frac{1}{2}\left(\frac{y_{tj}-v_{tj}}{\sigma_{tj}}\right)^{2}\right)\quad
j=J_{1}+1,...,J
\end{equation}
where $\sigma_{tj}^{2}$ is the variance of the continuous responses supposed
to be heteroscedastic over time and between items.

As in the classical generalized linear model, $\upsilon_{tj}$ is the linear predictor for
the $j$th outcome at time $t$ and the link between the linear predictor and the conditional means of the random distributions can be any monotonic differentiable function. In this context, the link is the logit of the probability associated to each count for the binomial distribution defined in eq. (\ref{binomial}) and the identity function in the case of the normal distribution defined in eq. (\ref{normal}).

For both kinds of observed variables the linear predictor is defined as
\begin{equation}\label{predictor}
\upsilon_{tj}=\sum_{r=0}^{p}\beta_{rj}\lambda_{t}^{r}+\sum_{k=1}^{q}\lambda_{kj}z_{k}+\sum_{l=1}^{b}\gamma_{tjl}x_{tjl}
\qquad t=1,...,T_{j}, \quad j=1,...,J
\end{equation}
and in matrix form
\begin{equation}
\upsilon_{tj}= \boldsymbol w_{tj} \boldsymbol \eta + \boldsymbol \gamma_{tj} \boldsymbol x_{tj}, \qquad j=1,...,J
\end{equation}
where
\begin{eqnarray} \nonumber
\nonumber \boldsymbol w_{t1}&=&[1,
\lambda_{t},...,\lambda_{t}^{p},...,0,...,0,\lambda_{11},
\lambda_{21},...,\lambda_{q1}],\\ \nonumber \cdots && \cdots \qquad \cdots\\
\nonumber \boldsymbol w_{tJ}&=&[0,...,0,...,1,
\lambda_{t},...,\lambda_{t}^{p},\lambda_{1J},
\lambda_{2J},...,\lambda_{qJ}],\\
\nonumber \boldsymbol \gamma_{tj}&=&[\gamma_{tj1},
...,\gamma_{tjb}],\\
\nonumber \boldsymbol x_{tj}&=&[x_{tj1},
...,x_{tjb}].
\end{eqnarray}

\noindent The latent variables $\boldsymbol
\eta=(\beta_{0j},...,\beta_{pj}, z_{1},...,z_{q})$ are random
effects and latent traits that account for both the temporal and the
cross-sectional dependence between items. The $\gamma_{tjl}$ are
fixed regression parameters representing the effects of the
covariates $x_{tjl}, l=1,...,b$. As it is done with "classical"
univariate growth models, the random coefficients \mbox{$\boldsymbol
\beta_{j}=(\beta_{0j}, ...,\ldots,\beta_{pj})$}, $ j=1,...,J$, and
the corresponding loadings $\lambda_{t}, t=1,...,T_{j}$, are
introduced in order to describe the temporal behaviour of each item,
$p$ being the degree of the fitted trajectory. The $\lambda_{t}$'s
either can be fixed, in the case of linear polynomials, or can be
parameters to be estimated if a nonlinear function is more
appropriate. The model is very flexible since it allows to specify
different temporal dynamics for each item. The common factors
$\boldsymbol z=(z_{1},\ldots ,z_{q})$ can represent traits of an
individual (\emph{e.g.} general and specific student abilities,
intelligence, etc. ) and determines the correlation between multiple
responses despite their temporal behaviour.

By defining the linear predictor in this way, the temporal
dependence between items as well as the autocorrelation of each item
are explained by variance and covariance elements related to the
random growth parameters $\boldsymbol \beta_{j},  j=1,...,J$. On the
other hand, the cross-correlation between items despite their
temporal behaviour is caught by the factor model via the loadings
$\boldsymbol \lambda_{jz}=(\lambda_{j1}, ...,\lambda_{jq}),
j=1,...,J$. These assumptions are contained in the prior density
function of the latent variables, $h(\boldsymbol \eta)$, supposed to
be  a multivariate normal density with mean vector $\boldsymbol
\mu_{\boldsymbol
\eta}=(\mu_{\beta_{01}},...,\mu_{\beta_{p1}},...,\mu_{\beta_{0J}},...,\mu_{\beta_{pJ}},0,...,0)$
and covariance matrix $\boldsymbol \Psi =\left(
                                                                                                                                            \begin{array}{cc}
                                                                                                                                              \boldsymbol \Psi_{\beta} & \mathbf{0} \\
                                                                                                                                               \mathbf{0} &  \boldsymbol \Psi_{z} \\
                                                                                                                                            \end{array}
                                                                                                                                          \right),$
where $\boldsymbol \Psi_{\beta} $ is the full covariance submatrix
related to the random effects $\boldsymbol \beta_{j}, j=1,..,J$, and
$\boldsymbol \Psi_{z}$ is the submatrix of the block matrix related
to the latent factors $\boldsymbol z$. We assume that the random
coefficients $\boldsymbol \beta_{j}$ and the factors $\boldsymbol z$
are independent. Furthermore, some constraints must be placed to
ensure identifiability of the model based on the observed data.
 In particular, as in classical latent variable models, there is indeterminacy related to the scales of the
 latent factors $\boldsymbol z$ \citep{Jo:69}. This indeterminacy can be eliminated by either setting $\lambda_{1j}=1$ or letting the variances of the factors be one, with at least one of the loadings constrained to be positive for each factor.

\noindent As a consequence of the structural specification of the model via $h(\boldsymbol \eta)$, the covariances between different responses, seeing in the scale provided by the link function, are given by
\begin{eqnarray} \nonumber
\text{Cov}(v_{tji},v_{t'ji})&=&\boldsymbol w_{tj\beta}\Psi_{\beta}
\boldsymbol w_{t'j\beta}'+ \boldsymbol w_{jz}\Psi_{\beta}
\boldsymbol w_{jz}'\\ \nonumber \text{Cov}(v_{tji},v_{tj'i})&=&
\boldsymbol w_{tj\beta}\Psi_{\beta} \boldsymbol w_{tj'\beta}'+
\boldsymbol w_{jz}\Psi_{\beta} \boldsymbol w_{j'z}'\\ \nonumber
\text{Cov}(v_{tji},v_{t'j'i})&=&\boldsymbol
w_{tj\beta}\Psi_{\beta} \boldsymbol w_{t'ji\beta}'+ \boldsymbol
w_{jz}\Psi_{\beta} \boldsymbol w_{j'z}'\\ \nonumber
\end{eqnarray} where $\boldsymbol w_{tj\beta}$ and $\boldsymbol w_{jz}, t=1,2,...,T_{j}, j=1,2,...,J$, indicate the coefficients in $\boldsymbol w_{tj}$ related to the random effects $\boldsymbol \beta_{j}, j=1,...,J,$ and the latent traits $\boldsymbol z$, respectively.

\subsection{Estimation}

The parameters of the model are estimated through the Maximum
Likelihood (ML) method via the EM algorithm, since the latent
variables $\boldsymbol \eta$ are unobserved. The EM starts with
initial values of the parameters. The algorithm consists of an
expectation and a maximization step. In the expectation step the
expected score function from the complete likelihood ($\boldsymbol
y, \boldsymbol
\eta| \boldsymbol x$) given the covariates %($\boldsymbol y,
%\boldsymbol x$)
is computed.  In the maximization step
updated parameter estimates are obtained from the equations derived
in the E-step. The whole procedure is repeated until convergence.

For a random sample of size $n$, it follows by eq. (\ref{model})
that the complete log-likelihood is written as:
\begin{eqnarray}\nonumber
L &=& \sum_{i=1}^{n}\log f(\boldsymbol y_{i}, \boldsymbol \eta_{i}|
\boldsymbol x_{i})\\\label{like} &=& \sum_{i=1}^{n}\left[\log
g(\boldsymbol y_{i}| \boldsymbol \eta_{i}, \boldsymbol x_{i}) + \log
h(\boldsymbol \eta_{i})\right]
\end{eqnarray}
where $g$ is the likelihood of the data conditional on the covariates,
 the latent variables and the random effects and $h$ is the common distribution function of the latent traits
 and the random effects. From eq. (\ref{like}), we see that the first component depends on both the factor loadings
 $ \boldsymbol \lambda_{jz}, j=1,...,J$ and the variance parameters $\boldsymbol \sigma_{j}^{2}=(\sigma_{1}^{2},...,\sigma_{T_{j}}^{2}), j=J_{1}+1,...,J,$ whereas the second component depends on $(\boldsymbol \mu_{\boldsymbol \eta}, \boldsymbol \Psi)$.

\subsubsection{Estimation of $\boldsymbol \mu_{\boldsymbol \eta}$ and $\boldsymbol \Psi$}

From the normality of $\boldsymbol \eta$, the second component of
the log-likelihood given in eq. (\ref{like}) (up to a constant) for
an individual $i$ is written as
\begin{equation}
\log h(\boldsymbol \eta_{i})= -\frac{1}{2} \ln \boldsymbol \Psi
-\frac{1}{2}(\boldsymbol \eta_{i} - \boldsymbol \mu_{\boldsymbol
\eta}) \boldsymbol \Psi^{-1} (\boldsymbol \eta_{i} - \boldsymbol
\mu_{\boldsymbol \eta})'.
\end{equation}
The expected score function needed for the EM implementation is
taken with respect to the posterior distribution of the latent
variables $h(\boldsymbol \eta_{i}| \boldsymbol y_{i}, \boldsymbol
x_{i})$. The expected score function for the parameter vector
$\boldsymbol \mu_{\eta}$ becomes
\begin{equation}\label{score}
ES_{i}(\boldsymbol \mu_{\eta})=\int S_{i}(\boldsymbol \mu_{\eta})
h(\boldsymbol \eta_{i}| \boldsymbol y_{i}, \boldsymbol x_{i}) d
\boldsymbol \eta_{i}, \qquad i=1,...,n
\end{equation}
where $$S_{i}(\boldsymbol \mu_{\eta})=\frac{\partial \log
h(\boldsymbol \eta_{i})}{\partial \boldsymbol
\mu_{\eta}}=\boldsymbol \Psi^{-1}(\boldsymbol \eta_{i} - \boldsymbol
\mu_{\boldsymbol \eta}).$$ Similarly, we obtain the score function
for $\boldsymbol \Psi$, that is,
$$S_{i}(\boldsymbol \Psi)=\frac{\partial \log h(\boldsymbol \eta_{i})}{\partial \boldsymbol \Psi}=-\frac{1}{2}\boldsymbol \Psi^{-1}-\frac{1}{2}\boldsymbol \Psi^{-1}(\boldsymbol \eta _{i}- \boldsymbol \mu_{\boldsymbol \eta})(\boldsymbol \eta_{i} - \boldsymbol \mu_{\boldsymbol \eta})'\boldsymbol \Psi^{-1}.$$

By solving $\sum_{i=1}^{n}ES_{i}(\boldsymbol \mu_{\eta})=0$ and
$\sum_{i=1}^{n}ES_{i}(\boldsymbol \Psi)=0$ we get explicit solutions
for the maximum likelihood estimators of $\boldsymbol \mu_{\eta}$
and $\boldsymbol \Psi$.

\subsubsection{Estimation of $\boldsymbol \lambda_{jz}$ and  $\boldsymbol \sigma_{j}^{2}$}

The estimation of parameters  $\boldsymbol \lambda_{jz}, j=1,...,J$ and  $\boldsymbol \sigma_{j}^{2}, j=J_{1}+1,...,J$ depends on the first component of the log-likelihood given in eq. (\ref{like}). Under the conditional independence assumption, the log-likelihood of the count and continuous data can be written as
\begin{eqnarray}\nonumber
\log g(\boldsymbol y_{i}| \boldsymbol \eta_{i}, \boldsymbol
x_{i})&=&\sum_{j=1}^{J_{1}}\sum_{t=1}^{T_{j}}\left[\log \left(
                                                                     \begin{array}{c}
                                                                       n_{t} \\
                                                                       y_{tji} \\
                                                                     \end{array}
                                                                   \right)+y_{tji}\upsilon_{tji}-n_{t}\log\left(1+\exp(\upsilon_{tji})\right)
\right]+\\ \label{likecond} &+&
\sum_{j=J_{1}+1}^{J}\sum_{t=1}^{T_{j}}\left[-\frac{1}{2}\log(2\pi)-\frac{1}{2}\log(\sigma_{tj}^{2})-\frac{(y_{tji}-\upsilon_{tji})^{2}}{2\sigma_{tj}^{2}}\right]
\end{eqnarray}
The first component refers to count variables and will be used to
derive estimates of the factor loadings corresponding to such
variables, that is, $\boldsymbol \lambda_{jz},j=1,...,J_{1}$. The
expected score function of the parameter vector $\lambda_{jz}$ is
again taken with respect to the posterior $h(\boldsymbol
\eta_{i}|\boldsymbol y_{i}, \boldsymbol x_{i})$:
\begin{equation}\label{scorbin}
ES_{i}(\boldsymbol \lambda_{jz})=\int S_{i}(\boldsymbol
\lambda_{jz})h(\boldsymbol \eta_{i}|\boldsymbol y_{i}, \boldsymbol
x_{i})d \boldsymbol \eta_{i}, \quad i=1,...,n
\end{equation}
where
$$S_{i}(\boldsymbol \lambda_{jz})=\frac{\partial \log g( y_{tji}|\boldsymbol \eta_{i}, \boldsymbol x_{i})}{\partial \boldsymbol
\lambda_{jz}},$$ and
\begin{equation}\label{espbin}
\frac{\partial \log g(y_{tji}|\boldsymbol \eta_{i}, \boldsymbol
x_{i})}{\partial \boldsymbol
\lambda_{jz}}=\sum_{t=1}^{T_{j}}\boldsymbol
z_{i}\left(y_{tji}-n_{t}\frac{\exp(\upsilon_{tji})}{(1+\exp(\upsilon_{tji}))}\right)
\qquad j=1,...,J_{1}.
\end{equation}
By replacing eq. (\ref{espbin}) into eq. (\ref{scorbin}) and solving
$\sum_{i=1}^{n}ES_{i}(\boldsymbol \lambda_{jz})=0$ we get
non-explicit solutions for the parameter vector $\boldsymbol
\lambda_{jz}$. A Newton-Raphson algorithm is used to solve the
nonlinear maximum likelihood equations.

From the second component in the likelihood (\ref{likecond}) we
estimate factor loadings and variance components corresponding to
continuous items. The expected score functions for the parameters
$\boldsymbol \lambda_{jz}, j=J_{1}+1,...,J$, are given by
\begin{equation}\label{scornorm}
ES_{i}(\boldsymbol \lambda_{jz})=\int S_{i}(\boldsymbol
\lambda_{jz})h(\boldsymbol \eta_{i}|\boldsymbol y_{i}, \boldsymbol
x_{i})d \boldsymbol \eta_{i}, \qquad  i=1,...,n
\end{equation}
where
$$S_{i}(\boldsymbol \lambda_{jz})=\sum_{t=1}^{T_{j}}\boldsymbol z_{i}^{2} \lambda_{jzi}-\sum_{t=1}^{T} \boldsymbol z_{i} (y_{tji}-\sum_{r=0}^{p}\beta_{rj}\lambda_{t}^{r}).$$
Similarly, we obtain the score function for each element in
$\boldsymbol \sigma_{j}^{2}, j=J_{1}+1,...,J$, that is,
$$S_{i}(\sigma_{tj}^{2})=\sigma_{tj}^{2}-(y_{tji}-\upsilon_{tji})^{2} \qquad  t=1,...,T_{j}, \quad i=1,...,n.$$
By solving $\sum_{i=1}^{n}ES_{i}(\boldsymbol \lambda_{jz})=0,
\sum_{i=1}^{n}ES_{i}(\boldsymbol \sigma_{j}^{2})=0$ we get explicit
solutions for the maximum likelihood estimators of $\boldsymbol
\lambda_{jz}$ and $\boldsymbol \sigma_{j}^{2}$ for
$j=J_{1}+1,...,J$.

Integrals are approximated by using Gauss-Hermite quadrature points.
In order to apply the Gauss-Hermite approximation to the integral of
equations (\ref{score}), (\ref{scorbin}), and (\ref{scornorm}) we
consider the Cholesky decomposition of the covariance matrix
$\boldsymbol \Psi$ given by $\boldsymbol \Psi =\boldsymbol C
\boldsymbol C'$. As shown by \cite{CaMo:09}, this is necessary
because the non null submatrices of $\boldsymbol \Psi$, namely
$\boldsymbol \Psi_{\boldsymbol \beta}$ and $\boldsymbol
\Psi_{\boldsymbol z}$, are not diagonal.

The steps of the EM algorithm are defined as follows:
\begin{description}
  \item[Step 1:] choose initial estimates for the model parameters. Starting values for the loadings are obtained by fitting separate confirmatory factor analysis models at each time points.
  Initial values for the other parameters are chosen arbitrarily.
  \item[Step 2:] compute the expected score functions for all the parameters (E-step).
  \item[Step 3:] obtain improved estimates for the parameters by solving the nonlinear maximum likelihood equations for the parameters corresponding to the count items  and explicit solutions for the parameters of the continuous items and the latent distribution (M step).
  \item[Step 4:] repeat steps 2-3 until convergence is attained.
\end{description}

\section{Results}
We start the analysis by estimating a model for the overall dataset
of students observed at the three different time points. As already
discussed, the aim of the analysis is twofold: (1) analyze the
student careers over time with respect to the Number of Exams taken
in each occasion (NE) and the corresponding Average Marks (AM), and
(2) measure the general latent capability of the students.
Therefore, we analyze how the variables NE and AM change over time
by means of multivariate latent growth models, and we extend these
models by including a common factor that can explain the atemporal
variability that exists between the two items. In particular, since
only three different academic years are considered, a linear
polynomial model ($p=1$) could be appropriate to describe the
temporal pattern of both NE and AM. Measurement invariance over time
of the loadings in the one-factor model ($q=1$) is assumed. Thus,
the estimated model (denoted as Model A) is characterized by the
following linear predictor
\begin{equation*}
\upsilon_{tj}=\beta_{0j}+\beta_{1j}(t-1)+\lambda_{j}z = \boldsymbol w_{tj} \boldsymbol \eta, \qquad
t=1,2,3, \quad j=NE,AM
\end{equation*}
where
\begin{eqnarray} \nonumber
\nonumber \boldsymbol w_{tNE}&=&[1,0,1,2,0,0,0,0,\lambda_{NE}]\\
\nonumber \boldsymbol w_{tAM}&=&[0,0,0,0,1,0,1,2,\lambda_{AM}],\\
\nonumber \boldsymbol \eta&=&[\beta_{0NE},\beta_{1NE},\beta_{0AM},\beta_{1AM},z].
\end{eqnarray}
Binomial-logistic and Normal heteroscedastic linear regression models are estimated for the NE and AM, respectively.
The multivariate normal density of the latent variables $h(\boldsymbol \eta$) has mean vector  $\boldsymbol \mu_{\boldsymbol \eta}=(\mu_{\beta_{0NE}},\mu_{\beta_{1NE}},\mu_{\beta_{0AM}},\mu_{\beta_{1AM}},0)$ and covariance matrix $\boldsymbol \Psi =\left(
                                                                                                                                            \begin{array}{cc}
                                                                                                                                              \boldsymbol \Psi_{\beta} & \mathbf{0} \\
                                                                                                                                               \mathbf{0} &  1 \\
                                                                                                                                            \end{array}
                                                                                                                                          \right).$
For identification reasons, the variance of the common latent factor $z$ is set equal to 1.\\
FORTRAN and R codes have been implemented to estimate the model.
(They are available from the authors upon request). Parameter
estimates of Model A for the overall dataset are reported in Table
\ref{Tab:4}.

\begin{table}
\caption{\label{Tab:4} Estimates for the overall data set (standard
errors in brackets)} \centering \fbox{\begin{footnotesize}
\begin{tabular}{ll}
\textbf{Model A} %\hline
&\\\emph{Coefficients} & \emph{Estimates}\\\hline
\emph{Multivariate growth model}\\
$\hat{\mu}_{\beta_{0}NE}$&$\hspace{0.25cm} 0.249$ $(0.047)$\\
$\hat{\mu}_{\beta_{1}NE}$&$-0.443$ $(0.026)$\\
$\hat{\mu}_{\beta_{0}AM}$&$\hspace{0.25cm}23.98$ $(0.326)$\\
$\hat{\mu}_{\beta_{1}AM}$&$-0.113$ $(0.244)$\\
%\\
$\hat{\boldsymbol \Psi}_{\beta}$&$\left(
                           \begin{array}{cccc}
                             $\hspace{0.1cm}0.231$^{*} & &  & \\
                             $-0.107$^{*} & $\hspace{0.1cm}0.093$^{*} & & \\
                             $\hspace{0.1cm}1.004$^{*}& $-0.531$^{*}  &$\hspace{0.05cm} 4.980$^{*}& \\
                             $-0.363$^{*} & $\hspace{0.05cm} 0.413$^{*} & $-2.183$^{*} & $2.164$^{*} \\
                           \end{array}
                         \right)$\\
                        &\tiny{ *: significant at $5\%$ level.}\\
\emph{Factor  model}\\
$\hat{\lambda}_{NE}$ &$\hspace{0.15cm} 0.524$ $(0.073)$ \\
$\hat{\lambda}_{AM}$ &$\hspace{0.15cm} 2.581$ $(0.449)$ \\
$\hat{\sigma}_{1AM}^{2}$&$10.080$ $(0.340)$  \\
$\hat{\sigma}_{2AM}^{2}$&$\hspace{0.15cm}9.953$ $(0.176)$ \\
$\hat{\sigma}_{3AM}^{2}$&$14.568$ $(0.395)$\\
\end{tabular}
\end{footnotesize}}
\end{table}
It can be noticed that both NE and AM present similar temporal
dynamics, even if expressed on different scales. In terms of the
population mean trajectory,  NE presents a mean initial status equal
to 0.249, indicating that in the first year the students take, in
mean, around 4.5 exams, as expressed in the original scale. The
students' progress is described by the slope mean parameter
$\mu_{\beta1}$, equal to -0.443, reflecting the linear,
term-by-term, worsening in mean achievement during the second and
third years. Both the mean initial status and rate of growth are
coherent with the descriptive analyses we performed in Section 2,
since on average the number of exams, as derived by Table
\ref{Tab:2}, were 4.76, 4.43, and 4.47 in $t_{1}, t_{2}, t_{3}$,
respectively. Similarly, at the initial status, students obtain an
average mark (in mean) around 23.98, but this mean worsens over time
as indicated by the mean slope parameter $\hat{\mu}_{\beta_{1}AM}$
equal to $-0.133$.\\
By looking at the covariances specific of each item in
$\hat{\boldsymbol \Psi}_{\beta}$,  students present a higher
variability in the initial status than in the rate of growth, with a
negative correlation between initial status and slope, for both NE
and AM. Multivariate latent curves also allow to analyze the
covariation between the temporal dynamics of NE and AM by estimating
cross-covariances between random intercepts and slopes of the two
curves. There are positive and significant covariances between
$\beta_{0NE}$ and $\beta_{0AM}$ as well as between the random
slopes, $\beta_{1NE}$ and $\beta_{1AM}$, indicating that students
with higher (smaller) average marks in the first year tend to take a
higher (smaller) number of exams at $t_{1}$, and that students with
positive (negative) slopes for AM generally present a similar
pattern for NE. On the other hand, negative covariances are
estimated between $\beta_{0NE}$ and $\beta_{1AM}$ as well as between
$\beta_{0AM}$ and $\beta_{1NE}$. Differently from classical
multivariate latent growth modeling, these cross-covariances between
NE and AM curves are free from the effect of a common latent factor
$z$ we estimated via integrating the growth curves with a one factor
model.

When manifest variables are of different types, care is needed in
the interpretation of the factor loadings, depending on the scale of
the $y_{tj}$. In order to interpret the latent factor $z$ we shall
therefore have to ensure that the $\lambda$s are calibrated so that
they may be meaningfully compared across variable types. This may be
done in a variety of ways but we follow the approach of
\cite{TaLe:87} and \cite{BaKn:99}, which provides a parametrization
that keeps the interpretation as close as possible to the familiar
methods of traditional factor analysis. This approach is based on a
standardization of the coefficients of
the latent variable $z$ in order to express correlation coefficients between the manifest variable $y_{tj}$ and
the  factor $z$.\\
For the normal item, $\lambda_{AM}$ denotes the covariance between
the manifest variables $y_{tAM}$ and the factor $z$. By dividing
$\lambda_{AM}$ by the square root of the variance of the continuous
variable $y_{tAM}, t=1,2,3$, we obtain the correlation between
$y_{tAM}$ and $z$,  that is
\begin{equation} \label{Stnor}
\lambda_{tAM}^{*}=\frac{\lambda_{AM}}{\sqrt{\psi_{\beta_{0AM}}^{2}+2(t-1)\psi_{\beta_{0AM},\beta_{1AM}}+(t-1)^{2}\psi_{\beta_{1AM}}^{2}+\lambda_{AM}^{2}+\sigma_{tAM}^{2}}}, \qquad t=1,2,3.
\end{equation}
Notice that the correlation varies over time, hence
\begin{eqnarray*}
\hat{\lambda}_{1AM}^{*}&=&0.554 \\
\hat{\lambda}_{2AM}^{*}&=&0.586 \\
\hat{\lambda}_{3AM}^{*}&=&0.505
\end{eqnarray*}
The amplitude of the factor loadings is quite similar in all the three occasions and on average is equal to $0.548$, indicating that the measurement invariance assumption is appropriate.\\
On the other hand, for the binomial item, the standardization
follows that proposed for binary items \citep{MoKn:01} and based on
the equivalence of the response function and underlying variable
approaches \citep{TaLe:87}.
 In this context, the correlation between a normal variable supposed to be underlying the binomial discrete observations and the latent variable $z$ is given by
\begin{equation}\label{Stbin}
\lambda_{tNE}^{*}=\frac{\lambda_{NE}}{\sqrt{\psi_{\beta_{0NE}}^{2}+2(t-1)\psi_{\beta_{0NE},\beta_{1NE}}+(t-1)^{2}\psi_{\beta_{1NE}}^{2}+\lambda_{NE}^{2}+1}}, \qquad t=1,2,3.
\end{equation}
The estimated standardized binomial loadings are
\begin{eqnarray*}
\hat{\lambda}_{1NE}^{*}&=&0.427 \\
\hat{\lambda}_{2NE}^{*}&=&0.445 \\
\hat{\lambda}_{3NE}^{*}&=&0.435
\end{eqnarray*}
which are really close to each other; the amplitude of the
standardized factor loadings is on average $0.436$. The standardized
coefficients given for normal and binomial variables can be used for
a unified interpretation of the loadings, bringing the
interpretation close to factor analysis. The common factor $z$
explains the interrelationships between the two observed items net
from their temporal dependence. Both variables are significant
indicators of this latent capability. Moreover, they both influence
positively this unobserved construct. Ignoring the presence of a
common factor in multivariate latent growth models can lead to an
overestimation of the cross-variation among multiple curves.

The goodness of fit of Model A has been checked separately for the
count and continuous part \citep{MoKn:01}. As for the count part of
the model, significant information concerning the goodness of fit
can be found in the margins. In particular, the one-way margins of
the differences between the observed ($O$) and expected ($E$)
frequencies under the model are investigated; any large
discrepancies will suggest that the model does not fit well these
counts. The Chi square test as well as high-way margins are not
appropriate because of the sparseness of the data \citep{Rei:96}.
Table \ref{Tab:5} gives the GF-fit measures, calculated as
$((O-E)^{2}/E)$, for each Binomial variable \citep{BaSt:02}.
\begin{table}
\caption{\label{Tab:5} Count items: GF-fit values for the one-way margins.}
\centering
\fbox{\begin{footnotesize}
\begin{tabular}{lccc}
\emph{Counts} & $y_{1EN}$ &  $y_{2EN}$ &  $y_{3EN}$\\\hline
0&11.58&11.65&11.23\\
1&0.10&1.59&3.05\\
2&5.78&0.22&10.39\\
3&5.45&0.39&15.20\\
4&8.73&4.97&5.91\\
5&1.10&16.18&0.95\\
6&7.58&5.12&34.35\\
7&24.47&0.33&37.70\\
8&3.65&23.75&2.33 \\
9&-&31.37&0.04   \\
10&-&13.06&0.66   \\
11&-&0.31&7.94    \\
12&-&0.02&0 \\
13&-&-&0.41 \\
14&-&-&1.23
\end{tabular}
\end{footnotesize}}
\end{table}
We can observe that the GF-fits are not good, especially those on count 7 for $y_{1NE}$, on 8 and 9 for $y_{2NE}$, and
on counts 6 and 7 for $y_{3NE}$. Reasons of this misfitting of Model A on the overall sample will be next investigated.

For the normal part of the model we check the discrepancies between the sample correlation matrix and the one estimated from the model, as illustrated in Table \ref{Tab:6}
for the variables $y_{1AM},y_{2AM}$, and $y_{3AM}$.
\begin{table}
\caption{\label{Tab:6} Normal items: discrepancies between sample and estimated correlation matrices.}
\centering
\fbox{\begin{footnotesize}
\begin{tabular}{lccc}
\ & $y_{1EN}$ &  $y_{2EN}$ &  $y_{3EN}$\\\hline
$y_{1EN}$&0.00&-0.05&0.02\\
$y_{2EN}$&-0.05&0.00&-0.05\\
 $y_{3EN}$&0.02&-0.05&0.00\\
\hline
\end{tabular}
\end{footnotesize}}
\end{table}
The discrepancies  between observed correlations and those estimated
are particularly small, indicating that the fit of the model for the
normal variables is good.

In Section 2, we showed that among the 821 students, two different
temporal patterns were evident, one related to those students who
graduated at $t_{3}$ (Grad) and the other to students who did not
get the degree regularly at the third year (Nograd). Hence, we shall
analyze these two different groups of students in order to
investigate the reasons of the poorness of fit for the count part of
the Model A in the overall dataset. Therefore, in the following, we
fit Model A to the Grad and Nograd students, separately.

\subsection{The Graduate students}
We first consider the $195$ students who graduated at $t_{3}$, and
we start by fitting the Model A described above. The results of the
estimation are reported in Table \ref{Tab:7}. It can be noticed that
parameter estimates corresponding to both the multivariate latent
growth and the factor parts of Model A differ substantially from
what we obtained for the overall sample. As for the former, both NE
and AM present higher mean initial status than the overall sample.
They are equal to $6.05$ for NE and to $26.17$ for AM, as expressed
in the original scales. Furthermore, the mean trajectory for NE has
a worsening pattern over time, whereas the one corresponding to AM
presents an increasing but not significant temporal behaviour. By
looking at the variability around the mean trajectories, this is
significant in the initial status ($0.133$) and in the rate of
growth ($0.072$) corresponding to NE which also shows a negative
correlation between the random intercept and slope. On the other
hand, there is not a significant variability for AM with respect to
its mean trajectory, indicating that the Grad students show a
similar pattern over time. This finding is also evident in all the
(not significant) covariances between random coefficients of NE and
AM.

As for the factor part, the loading associated to the binomial
variable is very close to 0 and not significant, suggesting that the
number of exams for these students is not a measure of their
capability. However, also the loading associated to AM is not
significant. This means that for Grad students it makes no sense to
specify a common factor related to the variables AM and NE. A
justification could be found from the not significant association
between AM and NE for Grad in time $t_{2}$,
 as shown in Section 2. This evidence can be a hint to test the assumption of measurement invariance over time.
 If we estimate a model where this assumption is relaxed,
denote it as Model B (Table \ref{Tab:7}), we can notice that the
time dependent loadings are very different. This is particularly
true for the variable NE, for which the loading does not change
greatly in the first two time points but becomes negative in
$t_{3}$. For the variable AM the loading increases over time.
Clearly, the measurement invariance cannot be assumed. This result
is also confirmed by the AIC and BIC criteria that show that Model B
is better than Model A. However, from our viewpoint, Model B is
meaningless in the factor part. Moreover, if we look at the one-way
margins associated to Model B (Table \ref{Tab:8}) we can see that
again there are goodness of fit problems at time points $t_{2}$ and
$t_{3}$. \\These results for Grad highlight two different aspects of
the analysis. First of all, we have a slight individual variability
around NE and AM mean trajectories, indicating a similar temporal
behaviour of these students. Moreover, the higher values of these
means compared to those obtained for the overall sample show a good
performance of this group. Secondly, in this case the two variables
are not measures of a latent construct that in the overall sample we
identified as capability.
\begin{table}
\caption{\label{Tab:7} Estimates for graduate (Grad) students} %\hspace{-2.5cm}
\centering
\fbox{\begin{footnotesize}
\begin{tabular}{ll}
\textbf{Model A}   %\hline
&\\\emph{Coefficients} & \emph{Estimates}\\\hline
\emph{Multivariate growth model}\\
$\hat{\mu}_{\beta_{0}NE}^{2}$&$1.133$ $(0.086)$\\
$\hat{\mu}_{\beta_{1}NE}^{2}$&$-0.577$ $(0.050)$\\
$\hat{\mu}_{\beta_{0}AM}^{2}$&$26.172$ $(0.149)$\\
$\hat{\mu}_{\beta_{1}AM}^{2}$&$0.080$ $(0.061)$\\
\\
$\boldsymbol \Psi_{\beta}$ & $\left(
                       \begin{array}{cccc}
                         $0.133$^{*} & &  & \\
                         $-0.097$^{*} & $0.072$^{*} & & \\
                         $0.441$ & $-0.327$ & $2.197$ & \\
                         $0.044$ & $-0.026$ & $-0.086$ & $0.376$ \\
                       \end{array}
                     \right)$\\
                        &\tiny{ *: significant at $5\%$ level.}\\
\emph{Factor  model}\\
$\hat{\lambda}_{NE}$ &$0.005$ $(0.262)$ \\
$\hat{\lambda}_{AM}$ &$0.670$ $(3.360)$ \\
$\hat{\sigma}_{1AM}^{2}$&$0.702$ $(0.147)$\\
$\hat{\sigma}_{2AM}^{2}$&$1.292$ $(0.113)$\\
$\hat{\sigma}_{3AM}^{2}$&$0.328$ $(0.185)$\\
&\\ %\hline
AIC= 4304.090\\
BIC= 4309.602\\
\hline \hline
\\
\textbf{Model B}   %\hline
&\\\emph{Coefficients} & \emph{Estimates}\\\hline
\emph{Multivariate growth model}\\
$\hat{\mu}_{\beta_{0}NE}^{2}$&$1.193$ $(0.093)$\\
$\hat{\mu}_{\beta_{1}NE}^{2}$&$-0.594$ $(0.066)$\\
$\hat{\mu}_{\beta_{0}AM}^{2}$&$26.225$ $(0.192)$\\
$\hat{\mu}_{\beta_{1}AM}^{2}$&$0.188$ $(0.079)$\\
\\
$\boldsymbol \Psi_{\beta}$ & $\left(
                           \begin{array}{cccc}
                             $0.136$^{*} & &  & \\
                             $-0.103$^{*} & $0.078$^{*} & & \\
                             $0.445$ & $-0.340$ & $2.727$ & \\
                             $0.025$ & $-0.021$ & $-0.167$ & $0.317$ \\
                           \end{array}
                         \right)$\\
                        &\tiny{ *: significant at $5\%$ level.}\\
\emph{Factor  model}\\
$\hat{\lambda}_{1NE}$ &$0.300$ $(0.413)$ \\
$\hat{\lambda}_{2NE}$ &$0.465$ $(0.594)$ \\
$\hat{\lambda}_{3NE}$ &$-0.103$ $(0.185)$\\
$\hat{\lambda}_{1AM}$ &$0.156$ $(0.339)$ \\
$\hat{\lambda}_{2AM}$ &$0.440$ $(0.571)$ \\
$\hat{\lambda}_{3AM}$ &$0.734$ $(0.886)$\\
$\hat{\sigma}_{1AM}^{2}$&$0.616$ $(0.129)$\\
$\hat{\sigma}_{2AM}^{2}$&$1.300$ $(0.114)$  \\
$\hat{\sigma}_{3AM}^{2}$&$0.351$ $(0.177)$    \\
&\\ %\hline
AIC= 4226.847\\
BIC= 4273.520\\
\end{tabular}
\end{footnotesize}}
\end{table}

\begin{table}
\caption{\label{Tab:8} Count items: GF-fit values for the one-way margins,  Model B.}
\centering
\fbox{\begin{footnotesize}
\begin{tabular}{lccc}
\emph{Counts} & $y_{1EN}$ &  $y_{2EN}$ &  $y_{3EN}$\\\hline
2& 0.34 & 0.05 &-\\
3& 4.65&   -   &-\\
4& 7.46 &0.03  &-\\
5& 4.20 &16.28 &18.16\\                      6& 1.37 &58.32 &11.10\\
7& 5.29 &13.48 &19.52\\
8& 0.86 &15.12 &0.00\\
9&-     &23.81 &0.40  \\
10&-    &14.41 &0.50  \\
11&-    &0.02  &2.35   \\
12&-    &1.38- &-\\
13&-    &-     &- \\
14&-    &-     &30.51\\
\end{tabular}
\end{footnotesize}}
\end{table}

\subsection{The Nograduate students}
Coherently with the previous analysis, we first estimated Model A
for Nograd students. The results are reported in Table \ref{Tab:9}.
The growth model shows results similar to the overall sample with a
higher variability for AM with respect to its mean trajectory.
\begin{table}
\caption{\label{Tab:9} Estimates for Undergraduate (Nograd) students} %\hspace{-2.5cm}
\centering
\fbox{\begin{footnotesize}
\begin{tabular}{ll}
\textbf{Model A} %\hline
&\\\emph{Coefficients} & \emph{Estimates}\\\hline
\emph{Multivariate growth model}\\
$\hat{\mu}_{\beta_{0}NE}^{2}$&$-0.030$ $(0.055)$\\
$\hat{\mu}_{\beta_{1}NE}^{2}$&$-0.353$ $(0.035)$\\
$\hat{\mu}_{\beta_{0}AM}^{2}$&$23.337$ $(0.431)$\\
$\hat{\mu}_{\beta_{1}AM}^{2}$&$-0.183$ $(0.326)$\\
\\
$\boldsymbol \Psi_{\beta}$ & $\left(
                           \begin{array}{cccc}
                             $\hspace{0.02cm} 0.171$^{*} & &  & \\
                             $-0.109$^{*} & $\hspace{0.02cm} 0.131$^{*} & & \\
                             $\hspace{0.02cm} 0.930$^{*} & $-0.610$^{*} & $ 5.514$^{*} & \\
                             $-0.552$^{*} & $\hspace{0.02cm} 0.658$^{*} & $-3.304$^{*} & $3.492$^{*} \\
                           \end{array}
                         \right)$\\
                        &\tiny{ *: significant at $5\%$ level.}\\
\emph{Factor  model}\\
$\hat{\lambda}_{NE}$ &$0.473$ $(0.119)$ \\
$\hat{\lambda}_{AM}$ &$2.609$ $(0.706)$ \\
$\hat{\sigma}_{1AM}^{2}$&$12.356$ $(0.567)$\\
$\hat{\sigma}_{2AM}^{2}$&$12.484$ $(0.269)$\\
$\hat{\sigma}_{3AM}^{2}$&$17.307$ $(0.612)$\\
&\\ %\hline
AIC= 18773.348\\
BIC= 18788.483\\
\hline \hline\\
\textbf{Model C}  %\hline
&\\\emph{Coefficients} & \emph{Estimates}\\\hline
\emph{Multivariate growth model}\\
$\hat{\mu}_{\beta_{0}NE}^{2}$&$-0.031$ $(0.051)$\\
$\hat{\mu}_{\beta_{1}NE}^{2}$&$-0.353$ $(0.032)$\\
$\hat{\mu}_{\beta_{0}AM}^{2}$&$ 23.346$ $(0.406)$\\
$\hat{\mu}_{\beta_{1}AM}^{2}$&$-0.215$ $(0.294)$\\
\\
$\boldsymbol \Psi_{\beta}$ & $\left(
                           \begin{array}{cccc}
                             $\hspace{0.02cm} 0.184$^{*} & &  & \\
                             $-0.117$^{*} & $\hspace{0.02cm} 0.133$^{*} & & \\
                             $\hspace{0.02cm} 0.935$^{*} & $-0.619$^{*} & $5.180$^{*} & \\
                             $-0.570$^{*} & $\hspace{0.02cm} 0.709$^{*} & $-3.241$^{*} & $4.052$^{*} \\
                           \end{array}
                         \right)$\\
                        &\tiny{ *: significant at $5\%$ level.}\\
\emph{Factor  model}\\
$\hat{\lambda}_{NE}$ &$0.464$ $(0.079)$\\
$\hat{\lambda}_{AM}$ &$2.502$ $(0.620)$\\
$\hat{\sigma}_{AM}^{2}$&$13.642$ $(0.234)$\\
&\\ %\hline
AIC= 18777.719\\
BIC= 18791.260\\
\end{tabular}
\end{footnotesize}}
\end{table}
\begin{table}
\caption{\label{Tab:10} Count items: GF-fit values for the one-way margins, Model A, Nograd.}
\centering
\fbox{\begin{footnotesize}
\begin{tabular}{lccc}
\emph{Counts} & $y_{1EN}$ &  $y_{2EN}$ &  $y_{3EN}$\\\hline
0&  7.69& 11.56   &  5.01     \\
1& 0.86 & 1.67    &  4.91     \\
2& 7.91 & 0.18    &  9.11     \\
3& 3.48 & 1.33    &  8.28     \\
4& 2.16 & 7.46    &  0.17     \\
5& 1.01 & 12.83   &  2.96     \\
6& 9.86 & 0.33    &  20.67     \\
7&12.61 & 7.89    &  8.47       \\
8&0.27  & 10.92   &  0.41      \\
9&  -   & 14.13   &  0.27      \\
10&-    & 5.54    &  3.57      \\
11&-    &  -      &  3.21     \\
12&-    & 0.35    &  -  \\
13&-    &  -      &  1.53    \\
\end{tabular}
\end{footnotesize}} \vspace{-0.35cm}
\end{table}
In this case the results related to the factor model are very
interesting. Differently from what we found for the Grad group, the
loadings are both significant and positively related to the latent
variable and indicate that, as in the overall data set, the factor
model is appropriate. Differently from what we found in all previous
analysis, the GF-fits of this model are satisfactory for all the
observed time points, as reported in Table \ref{Tab:10}. Thus the
count part of the model is well fitted by the binomial distribution.
Also for the normal part the fit is very good, the discrepancies
between observed and estimated correlations being very low (Table
\ref{Tab:11}). Therefore, Model A fits well the Nograd students
data.
 \begin{table} %\vspace{0.5cm}
\caption{\label{Tab:11} Normal items: discrepancies between sample and estimated correlation matrices, Model A, Nograd.}
\centering
\fbox{\begin{footnotesize}
\begin{tabular}{lccc}
\ & $y_{1EN}$ &  $y_{2EN}$ &  $y_{3EN}$\\\hline
$y_{1EN}$&0.00&-0.07&0.02\\
$y_{2EN}$&-0.07&0.00&-0.08\\
 $y_{3EN}$&0.02&-0.08&0.00\\
\hline
\end{tabular}
\end{footnotesize}}
\end{table}

If we look again at Table \ref{Tab:9} Model A, it can be noticed
that the values of $\sigma^{2}_{AM}$ are quite similar over time;
thus, it can be interesting to evaluate if AM is homoscedastic over
time. The results of the estimation of the model with homoscedastic
errors (Model C) are reported in the bottom part of Table
\ref{Tab:9}. Although all the parameter estimates do not change
abruptly, the AIC and BIC are slightly better for Model A than for
Model C, suggesting that such assumption does not hold. Thus, the
comparison between the loadings of AM and NE for evaluating the
influence of each item on the latent variable requires their
standardization according to eq. (\ref{Stnor}) for AM and to eq.
(\ref{Stbin}) for NE. We get the following standardized loadings:
$\hat{\boldsymbol \lambda}_{AM}^{*}=(0.525, 0.560, 0.534)$, and
 $\hat{\boldsymbol \lambda}_{NE}^{*}=(0.401,
0.414,0.388)$. As in the overall data set, the correlation between
$z$ and AM is slightly higher than that between $z$ and NE.

\section{Discussion}
In this paper we extended and applied multivariate latent growth
models to the analysis of student record data collected repeatedly
in the Data warehouse system of the University of Bologna. The
proposed approach is innovative since it allows to evaluate both the
student performance over time and individual capabilities
simultaneously. Key features include \textit{i)} a flexible modeling
of the temporal dynamics of the observed variables via specific
latent curves, and \textit{ii)} an extension of the multivariate
growth model that incorporates a factor part. Such component
explains the association between the observed items by means of
latent variables, interpreted as different traits or capabilities.\\
The complexity of the  model proposed lies in different aspects,
such as the presence of mixed data, the possibility of both
including several latent variables/random effects and estimating
specific temporal patterns for the observed variables. Hence,
computational problems occur in the parameter estimation. We
successfully solved them by implementing an ad hoc EM algorithm
(Fortran and R code are available upon request by the authors). As
far as we know, commercial software does not allow to treat all
these aspects simultaneously.

 We demonstrated, via different specifications of the model,
how our general approach can provide insights into the data
structure. In particular, the analysis carried out on a cohort of
students enrolled at the Faculty of Economics observed at three
different time points highlighted an heterogeneity in the overall
data set in terms of both average marks and number of exams. This is
due to the presence of different temporal patterns within the
cohort, since we have students who regularly graduate at $t_{3}$
(Grad) and students who
 did not manage to get the degree within the third year
(Nograd).  Grad students perform very well in terms of both number
of exams and average marks with similar temporal pattern. Nograd
take a lower number of exams with lower average marks, but within
this group we have a significant variability both in the initial
status and in the rate of growth. The factor part of the model for
the Grad student fails in measuring a general capability by means of
the observed indicators considered. We found that this fact depends
on the fundamental assumption of measurement invariance of items
over time. Such assumption
 does not hold in this case.
On the contrary, the model fits well the data of Nograd students.
What we called atemporal latent capability is well measured by the
average mark and the number of exams taken, both being significantly
related to the latent variable. The good performance of the model is
confirmed by the analysis of some goodness of fit statistics.\\ The
heterogeneity observed in the patterns of graduate and nograduate
students has implications on the results of the overall data set. On
the one hand parameter estimates for the overall sample are quite
similar to those of Nograd students, the factor loadings of the two
variables being both significant. This is in part due to the larger
sample size of Nograd students. However, the presence of different
performances of the Grad students reflects on the poorness of fit of
the count variable in the overall data set.

The model proposed here was motivated by the study of the students'
achievements and the good results obtained clearly show its
appropriateness. However, such methodology can be applied
successfully in many other fields, such as socio-economic settings
in which personal behaviours are studied by using panel data
collected through the administration
 of questionnaires.

In the present example no covariates have been considered. In
practice, we may have useful time dependent and time independent
covariates such as gender, region of origin, age, etc.  that can be
incorporated into the model. In particular,  an emerging field of
investigation is based on the comparison of the performances of
students who completed the degree compared with those who abandoned
\citep{SmNa:01,DrGi:04}. Furthermore, the treatment of missing data
would allow to extend our analysis to more time points and evaluate
if non linear or higher degree polynomial trajectories can describe
the temporal behaviour of the items studied. Preliminary studies
performed with the software LISREL \citep{BiCa:07}  showed how
different latent curves can fit the weighted average marks for
different groups of students within the cohort analyzed. Such
problems will motivate our future investigations along these lines
of research.

%\section*{Acknowledgements}

%\vspace{-1cm}

%\bibliographystyle{Chicago}

\end{document}